# Energy Saving and Traffic Steering Use Case and Testing by O-RAN RIC xApp/rApp Multi-vendor Interoperability

Arda Akman
Juniper Networks

Pablo Oliver
Vodafone

Michael Jones
Keysight Technologies

Peyman Tehrani
AirHop Communications

Marcin Hoffmann
Rimedo Labs

Jia Li
Keysight Technologies

*Abstract*—This paper discusses the use case of energy saving and traffic steering in O-RAN, the mechanism of multi-vendor interoperability to make it work and depict its test methodology.

## I. O-RAN Introduction

Open RAN enables service providers to unlock historically closed, proprietary radio access networks (RAN) to a growing ecosystem of third-party vendors to build an intelligent and smarter RAN through open interfaces [1]. The closed RAN data, which was previously accessible only to RAN vendors, is now available for service providers and third-party vendors to build innovative applications on top of RAN Intelligent Controller (RIC) platform. rApp sitting together with Non-RT RIC and xApp working with Near-RT RIC of different purpose can utilize these data from O-RAN E2 node to make Artificial Intelligent / Machine Learning (AI/ML) analysis and optimize the RAN resources usage and save cost for mobile network operators. Recently, Energy Efficiency (EE) has become one of the crucial directions for 5G and beyond developments [2]. The aim of this paper is to demonstrate the multi-vendor setup exploiting Traffic Steering xApp (TS-App) from Rimedo Labs, Energy Saving rApp (ES-rApp) from AirHop Communications deployed within the Juniper Networks RIC (xApp in Near-RT RIC, and rApp in Non-RT RIC). The applications are validated with the use of Keysight's RICtest empowered by the anonymized data provided by Vodafone.

The rest of the paper is organized as follows: Sec. II provides description of options for cooperation between Rimedo TS-xApp and third-party rApp. Sec. III describes the AirHop ES rApp together with its underlying algorithms. Sec. IV aims at detecting and managing conflicting interactions between xApps/rApps by Juniper RIC. Finally, Sec. V describes the test methodology of ES and TS use case using Keysight's RICtest and Vodafone data and provides the discussion on the obtained results.

## II. Background of Energy Saving and Traffic Steering

One of the possibilities to significantly improve the EE of mobile networks is to dynamically switch cells on/off depending on the network traffic conditions. However, to switch off certain cells, first, the users must be offloaded to the neighboring ones. To achieve this one should utilize a Traffic Steering (TS) algorithm. TS is one of the fundamental mechanisms in 5G networks that allows Mobile Network Operators (MNOs) to distribute network traffic between cells, e.g., to equalize load or separate Quality of Service (QoS) flows. As such, it has been defined by O-RAN ALLIANCE as one of the first use cases to be addressed by xApps [3]. The algorithms aiming to improve EE are usually implemented as rApps, while TS is typically realized as xApps, for non-real-time (RT) and near-RT control loops, respectively. The challenge is that the TS-xApp must be notified by the ES-rApp that there is an identified candidate cell for switch-off or switch-on, which changes the Admissions Control state of the candidate cell. Further, in case of switch-off or switch-on, TS xApp can participate with service to offload or onboard subscribers to the candidate cell while taking into account neighboring cell loads and users' flows.

In this section we discuss two options for notification of the TS-xApp on the Energy Saving (ES) actions from the O-RAN specifications perspective. The first option is to utilize the A1 interface to send policies with a "FORBID" clause for cells that will be switched off, to indicate TS-xApp that users should be removed from these cells. It was presented during O-RAN Global Plugfest Fall 2023 in the setup with Juniper Near-RT RAN Intelligent Controller (RIC) and Keysight RICtest. The second approach is to utilize the E2 interface with its Cell Configuration and Control Service Model (E2SM-CCC). This approach was presented during the RIC Forum in Dallas (26-28 March 2024) organized by the U.S. National Telecommunications and Information Administration (NTIA) and the U.S. Department of Defense (DoD) [12].

### A. Traffic Steering xApp notification through the A1 policy in the Energy Saving use case

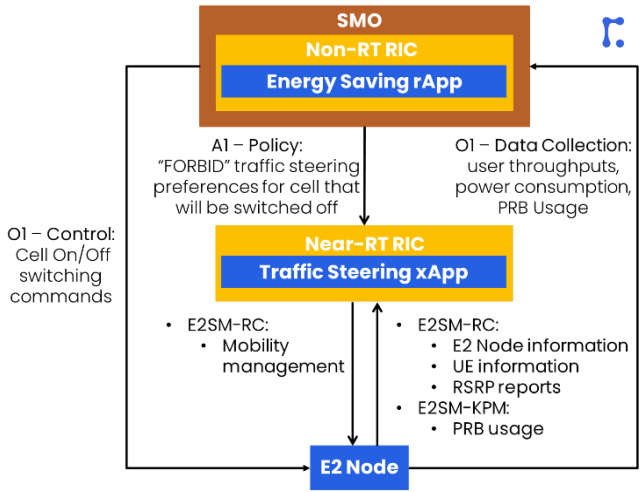

Fig. 1. Block diagram of TS xApp notification about the cell On/Off switching through the A1 policy (source [10])

In Fig. 1 we can see a setup with ES-rApp residing in the Non-RT RIC, and TS-xApp deployed in the Near-RT RIC. The ES-rApp utilizes the O1 interface to switch on/off cells based on the observed Key Performance Indicators (KPIs), e.g., user throughput, power consumption, and Physical Resource Block (PRB) usage. The TS-xApp subscribes to E2SM RAN Control (E2SM-RC) to obtain E2 Node, UE information, and measurement reports [4]. The xApp also obtains cell load (PRB usage) from E2SM KPM [5]. Based on this information, the xApp performs TS actions using E2SM-RC. In this setup, the A1 interface is used to provide TS-xApp with notifications about the cells that are subject to be switched off. This is done by formulating the TSP (Traffic Steering Preference) policy that contains a "FORBID" clause (see A1 interface specification for details [6]) for all cells that ES-rApp decides to switch off. After receiving such a policy, TS-xApp starts moving users to other cells. In the meantime, the ES rApp must monitor the number of RRC connections in these cells, and when they reach zero, it can put them into the ES mode. In the opposite direction, the ES-rApp switches the particular cell on, and once all the procedures are finished, it sends a request to delete the "FORBID" policy through the A1 interface, so that TS-xApp can utilize these cells. The evaluation of this procedure using Rimedo Labs TS-xApp deployed at the commercial Near-RT RIC and connected to RICtest software will be discussed next.

One of Rimedo Labs' activities during the O-RAN Global Plugfest Fall 2023 was the integration of the TS-xApp with Juniper's Near-RT RIC and Keysight RICtest as depicted in Fig. 2. The Rimedo TS-xApp was deployed within the Juniper Near-RT RIC as a docker container. The RIC was a broker between the xApp and Keysight RICtest, responsible for providing the E2 and A1 interface termination for both REPORT and CONTROL actions. The Keysight RICtest realistically emulated E2-Nodes with message exchange fully compliant with the O-RAN specifications. We used E2SM-RC (version 1.03) to:

- Retrieve the RSRP using: E2SM-RC REPORT Style 1 Message Copy

- Retrieve the Cell Global Identifier (CGI) and Physical Cell ID (PCI) using: E2SM-RC REPORT Style 3 E2 Node Information Change

- Retrieve the UE ID using: E2SM-RC REPORT Style 4 UE Information Change

- Control Handovers using: E2SM-RC CONTROL Style 3 Connected Mode Mobility

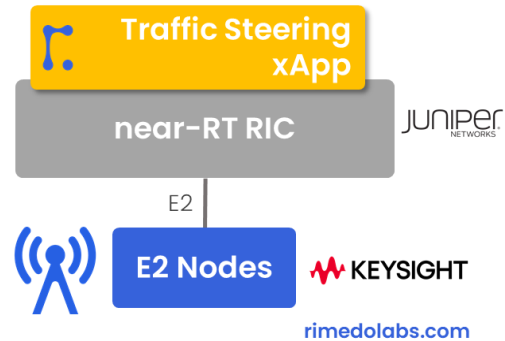

Fig. 2. Block diagram of the setup with Rimedo Labs TS-xApp, Juniper Near-RT RIC, and Keysight RICtest

One of the presented test cases was offloading the users from the cell based on the A1 policy, which fits the Traffic Steering xApp notification through the A1 policy described in this blog post [10]. The operation is presented in Figure 3. The left part of the figure presents the per-cell number of connected UEs (RRC connections) and handover statistics before the application of the A1 policy that is intended to offload users from Cell #2. We can see 10 UEs connected to Cell #2, and 2 UEs connected to Cell #4. The state of the system after application of the A1 policy with the 'FORBID' clause for Cell #2 is depicted on the right side of Fig. 3. We can see that UEs were moved from Cell #2 to Cell #4 which now serves 12 UEs. The green bar at the left side indicates 100% success of initiated handovers. Also, the number of handovers from Cell #2 is being updated and currently showing a value of 9. Finally, the number of sent control messages is increased to 21. This then should result in an O1 message sent to the ES-rApp stating that the cell is empty, and can be switched off.

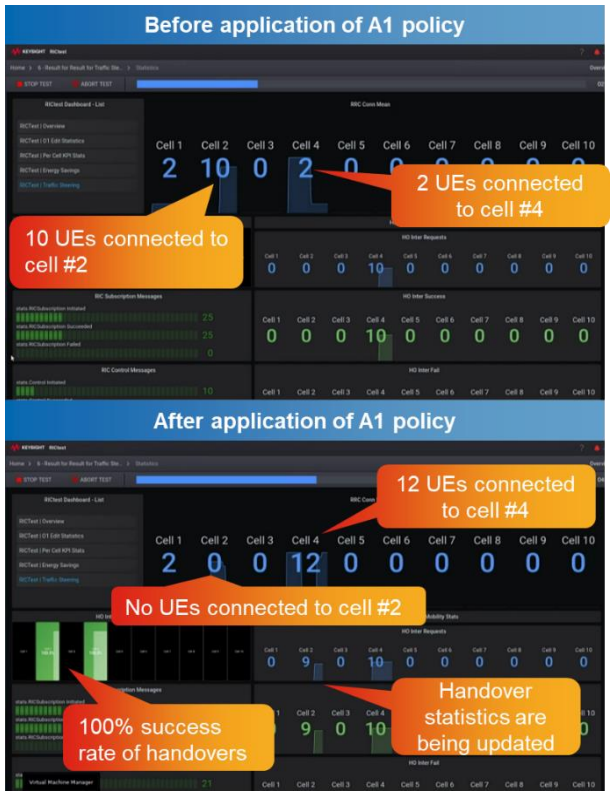

Fig. 3. Keysight RICtest dashboard view before and after the application of the A1 policy (based on [9])

## B. Traffic Steering xApp notification through E2SM-CCC in Energy Saving use case

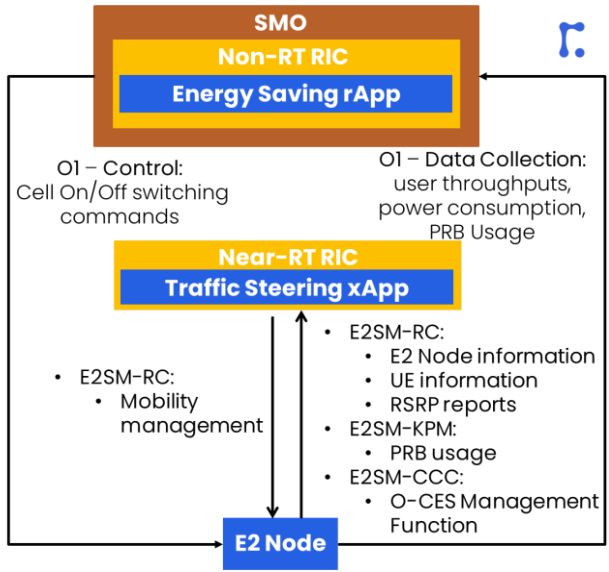

Fig. 4. Block diagram of TS-xApp notification about the cell on/off switching through the E2SM CCC (source [10])

The previous sections presented the concept of notification of TS-xApp about ES-rApp cell on/off decisions through the A1 interface. This section discusses a different approach that utilizes the E2 interface. The setup depicted in Figure 4 is similar to the one from Figure 1. The difference is that there is no A1 interface utilized. Instead, the TS xApp subscribes to E2SM-CCC to listen to the changes in the O-CES Management Function. E2SM-CCC allows xApps to both receive reports, of the configuration of a cell or E2-Node or to change this configuration using the so-called Configuration Structures originating from the 3GPP specifications. One such structure dedicated to ES functions is the already mentioned O-CES Management Function [7]. It contains parameters that inform an xApp about the ES features of a given cell [8].

- **cesSwitch** can be TRUE or FALSE to indicate if the ES features are enabled for a given cell.

- **energySavingState** specifies the ES state of the cell. It can be either "isNotEnergySaving" or "isEnergySaving". The "isNotEnergySaving" state refers to the standard operation of a cell, with all hardware components being active. The "isEnergySaving" state of the cell indicates that the cell is currently providing ES, and most of its hardware is deactivated. Such a cell cannot serve users.

- **energySavingControl** is used to either initialize ("toBeEnergySaving") or deactivate ("toBeNotEnergySaving") an ES state for a given cell.

The "toBeEnergySaving" refers to the transition state from "isNotEnergySaving" to "isEnergySaving," while the "toBeNotEnergySaving" refers to the reversed order transition. While utilizing the E2SM-CCC the TS-xApp receives an indication message whenever any of the O-CES Management Function attributes change, and based on this information, it can offload users from certain cells. In this approach, the ES-rApp does not have to be aware of the TS-xApp. It simply triggers cell deactivation, and TS-xApp receives proper notification through the E2 interface. Then, the TS-xApp moves users from cells that are in the "toBeEnergySaving" state. After all users are removed from the cell, its state can be simply changed to "isEnergySaving" by the E2-Node itself to enable significant ES. When the ES-rApp decides to switch on the cell, the TS xApp waits until the cell state is changed to "isNotEnergySaving" and starts to utilize it for receiving the user traffic.

The TS-xApp notification through E2SM-CCC has been introduced in Rimedo Labs TS-xApp and demonstrated in a multi-vendor setup during the RIC Forum in Dallas organized by the NTIA and U.S. DoD in March 2024. The setup included Rimedo TS-xApp deployed in Juniper Near-RT RIC, AirHop ES-rApp deployed in Juniper Non-RT RIC, and Keysight RICtest emulating the real 5G network and O-RAN interfaces based on the data provided by the Vodafone. The following sections of this paper will describe the design of ES-rApp, as well as the obtained results.

## III. AIRHOP ENERGY SAVING RAPP FOR MULTI-CARRIER RAN DEPOLYMNET

### A. Problem Statement

Communication Service Providers (CSPs) face the dual challenge of meeting growing wireless service demand while managing operational expenses (OpEx) and minimizing environmental impact. Continuous operation at maximum capacity leads to inefficiencies and high energy consumption. To address this, we introduce an intelligent energy savings application (rApp) deployed on a non-Real-Time RIC in the Open RAN architecture. This rApp uses AI-based learning and RAN programmability to optimize energy consumption in multi-carrier RANs by making intelligent decisions on when to switch capacity carriers OFF or ON, while maintaining user Quality-of-Experience (QoE). Variable service demands require a dynamic approach to energy management, which the rApp effectively provides.

There are several key challenges influencing the appropriate decisions for putting a capacity carrier into or out of a low energy state: Variable Demand, where service demand fluctuates significantly, requiring a flexible and adaptive energy management solution; Energy Consumption, as maintaining carriers during low utilization periods is inefficient, costly, and unsustainable; and User Quality-of-Experience (QoE), ensuring that any energy-saving measures do not compromise user QoE.

### B. Proposed solution

The proposed Energy Savings rApp leverages AI-based learning algorithms and RAN programmability to predict optimal times to switch RAN elements OFF/ON, minimizing energy consumption while ensuring user QoE. It differentiates between the coverage layer, which ensures basic connectivity, and the capacity layer, which handles high traffic and data-intensive applications, dynamically balancing these layers by activating additional capacity only when necessary.

The AI system continuously monitors network conditions and predicts load variations using historical data and ML algorithms. Key steps include real-time monitoring of traffic patterns and load distribution and load prediction to manage RAN element activation or deactivation. The AI model operates on load-based decision-making: turning off the capacity layer when the coverage layer can handle the load and turning it on when the coverage layer cannot sufficiently service the load to avoid QoE degradation.

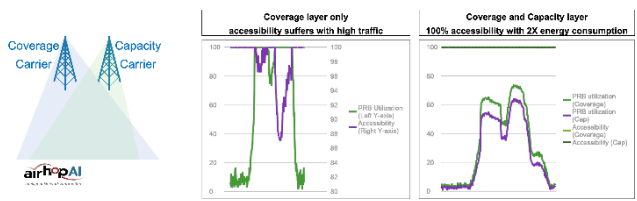

Fig. 5.  Comparison of load and accessibility between scenarios where only the coverage layer is on and where both coverage and capacity layers are active.

The AI model training leveraged a comprehensive commercial network dataset form Vodafone:

- Dataset Details: The training dataset comprises data from 13 sites and 41 sectors across 5 frequency bands, with a granularity of 15 minutes over 2 weeks.

- Metrics: Key metrics in the training consideration include physical resource block (PRB) utilization as an indicator of traffic load, power consumption, user QoE satisfaction (accessibility), the number of ON-OFF transitions, and model complexity.

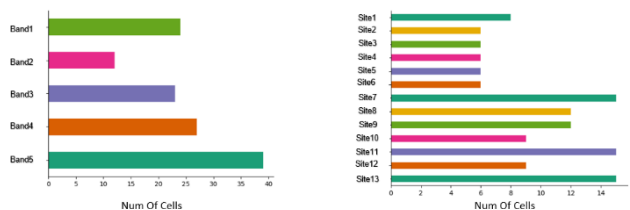

Fig. 6.  Number of cells across different frequency bands and sites in the Vodafone dataset.

### C. ESMC rApp Results

The performance of the Energy Savings for Multi-Carrier (ESMC) rApp was evaluated using the Keysight RICtest to simulate a representative network based on the Vodafone dataset. The results demonstrated significant improvements in energy efficiency while maintaining high levels of network accessibility. The ESMC rApp operates by dynamically switching OFF/ON capacity carriers in the network, resulting in different energy saving modes based on network load predictions as shown Fig. 7. Here energy saving mode indicates the number of capacity caries that are activated.

The implementation of the ESMC rApp resulted in approximately 25% energy consumption reduction achieved on the capacity layers by dynamically adjusting the number of active carriers in response to traffic load. One of the critical performance criteria for the ESMC rApp was maintaining high accessibility levels to ensure user QoS. The results indicate the system maintained an exceptional accessibility level of 99.999%, ensuring that the energy-saving measures did not negatively impact user experience.

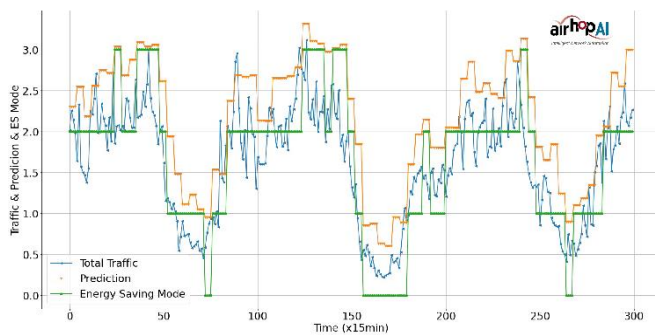

Fig. 7.  Traffic load, prediction, and energy-saving modes over time with AirHop's ESMC rApp.

The ESMC rApp intelligently managed RAN elements, resulting in significant energy reductions. During low traffic, the rApp minimized energy consumption by switching some or all capacity carriers to a low energy state. In high traffic, it balanced energy savings with performance by operating with minimal capacity carriers in a low energy state. The AI-driven approach ensured energy savings without compromising network performance, maintaining high user accessibility and quality assurance during mode transitions.

One important aspect of O-RAN is the ability to detect and manage conflicting interactions between multi-vendor rApps or xApps on RIC platform without any degradation of network performance, generally labeled as "conflict management". For example, Energy Savings and Traffic Steering use cases have very different goals and they optimize very different sets of parameters but could impact the functionality of each other.

An equally important opportunity is the coordination and cooperation between applications to provide service providers the flexibility to deliver new services and improved user experiences with greater agility and ease. For example, if the Energy Savings and Traffic Steering use cases cooperate with each other, they could jointly help save energy without any impact on the service quality. In the recent National Telecommunications and Information Administration (NTIA) RIC Forum, Juniper Networks, in partnership with Vodafone, AirHop, Rimedo Labs, and Keysight demonstrated multi-vendor rApp/xApp coordination with Energy Savings rApp and Traffic steering xApp, achieving 25% energy savings.

## IV. Test Methodology Of ES and TS Use Case

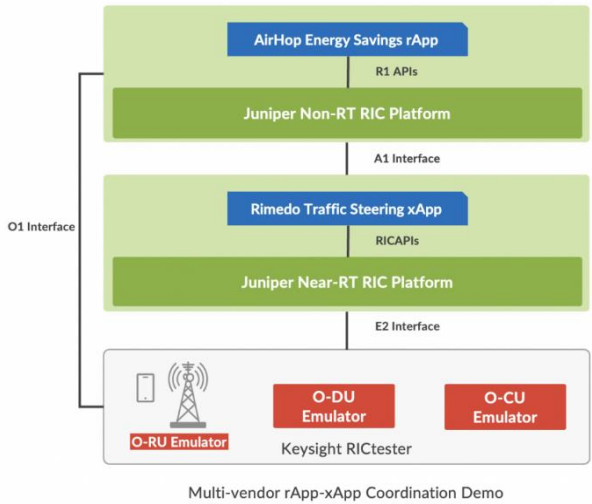

Fig. 8. Test topology of ES and TS use case

An important pre-requisite in replicating a real-world scenario was real network data and Vodafone provided access to an anonymized real network topology along with the performance data from this large commercial cluster. Keysight provided their RICtest solution to emulate the RAN environment based on the real-world network data provided by Vodafone. Keysight RICtest emulates UEs (User Equipment), cells, nodes, control/user plane events and supports O-RAN E2 and O1 interfaces.

As part of the test, Juniper provided its AI-Enabled Non-RT RIC and Near-RT RIC platforms, which were incubated in Juniper Beyond Labs and are now commercially available. Both the Juniper Non-RT RIC and the Near-RT RIC are based on a cloud-native microservices architecture and are fully compliant with O-RAN specifications and interfaces. They also support open APIs (R1 and RICAPI) for integration with any third-party O-RAN-compliant xApps or rApps, giving network operators greater flexibility and choice of suppliers. While the focus of this demo was the cooperation of multi-vendor applications through RIC platform services and O-RAN standards, Juniper RICs also provide conflict management and mitigation services to help prevent conflicts between applications.

AirHop implemented Dynamic Multi-Carrier Energy Savings Management rApp on top of Juniper Non-RT RIC platform. Service providers experience varied levels of service demand through the day and this rApp enables energy consumption to dynamically adapt to the mobile network service demand.

Typically, in any multi-carrier deployment, there is a coverage carrier with the ability to service a broad area of coverage at some level of capacity. There are also additional capacity carriers that can handle the traffic demand by balancing the traffic load across the different carriers. AirHop's Energy Savings rApp leverages AI-based learning and RAN programmability to predict when to turn RAN elements OFF/ON in the capacity carrier to minimize energy consumption. For example, the rApp can turn off one of the capacity carriers if the coverage carrier is not loaded. Also, if traffic starts to increase and coverage carrier becomes overloaded, the rApp can move the capacity carrier out of energy saving state to handle the increased traffic.

Traffic steering allows service providers to distribute network traffic between cells, for example, to equalize load or separate QoS flows. The traffic steering application is required to offload users to neighboring cells before any Energy Savings application can switch off the cells. Rimedo Labs implemented Traffic Steering xApp on top of Juniper Near-RT RIC platform. This xApp associates users with cells based on multiple factors, including radio conditions, cell load, cell types, service type/QoS profile, per-user, per-slice association, etc., and the cell energy status.

The traffic steering application supports both service-based traffic offloading and load balancing. For example, in a network with a highly loaded macrocell and underutilized small cell, with service-based offloading the mobile broadband users can be offloaded to the small cell and the voice users are served by macro cell. With load-balancing approach, the load can be shared between the macro cell and the small cell. AirHop Multi-Carrier Energy Savings Management rApp and Rimedo Labs Traffic Steering xApp work in tandem to ensure that energy savings are achieved without impacting the user QoS.

The test successfully showcased cooperative multi-vendor applications (AirHop's Energy Savings rApp & Rimedo Labs' Traffic Steering xApp) running on commercially available Non-RT RIC and Near-RT RIC platform from Juniper Networks. The interaction and the cooperation between AirHop's Energy Savings rApp and Rimedo Labs' Traffic Steering xApp was achieved with (O-RAN E2 Service Model Cell Configuration and Control (E2SM-CCC inherent design principles.

The energy savings actions requested by AirHop Energy Savings rApp toward the candidate cell simulated by Keysight RICtest triggered E2SM-CCC notifications towards Juniper Near-RT RIC/Rimedo's Traffic Steering xApp. Upon the receipt of this notification, Rimedo Labs' xApp started offloading the remaining users from this candidate energy saving cell to neighboring cells to minimize and eliminate any potential service impact. After all of the users were offloaded, the cell was set to energy savings state.

In terms of results, all the capacity carriers achieved energy savings of 25% with 99.999% accessibility based on the Vodafone anonymized dataset, as simulated by the Keysight RICtest. This is a notable example of how service providers can achieve energy savings while maintaining a premium experience for users.

## References

[1] M. Polese et al., "Empowering the 6G Cellular Architecture With Open RAN," in IEEE Journal on Selected Areas in Communications, vol. 42, no. 2, pp. 245-262, Feb. 2024, doi: 10.1109/JSAC.2023.3334610.

[2] M. Hoffmann, M. Dryjanski, "The O-RAN Whitepaper 2023 – Energy Efficiency in O-RAN", 2023, https://rimedolabs.com/blog/the-oran-whitepaper-2023-energy-efficiency-in-oran/

[3] O-RAN ALLIANCE, "O-RAN Working Group 1 Use Cases Analysis Report" v13.00, February 202

[4] O-RAN ALLIANCE, "O-RAN Working Group 3 Near-Real-time RAN Intelligent Controller E2 Service Model (E2SM), RAN Control" v01.03, October 2021

[5] O-RAN ALLIANCE, "O-RAN Working Group 3 Near-Real-time RAN Intelligent Controller E2 Service Model (E2SM), KPM", v02.0, October 2021

[6] O-RAN ALLIANCE, "A1 interface: Type Definitions", v07.00, February 2024

[7] O-RAN ALLIANCE, "O-RAN Work Group 3 (WG-3) Near-Real-Time RAN Intelligent Controller E2 Service Model (E2SM) Cell Configuration and Control", February 2024

[8] 3GPP, "TS 28.541; 3rd Generation Partnership Project; Technical Specification Group Services and System Aspects; Management and orchestration; 5G Network Resource Model (NRM); Stage 2 and stage 3 (Release 18)", v18.2.0, March 2024

[9] O-RAN ALLIANCE, "Verification of Rimedo Labs Traffic Steering xApp and Juniper Near-RT RIC using Keysight RICTest", Plugfest Fall, November 2023, https://plugfestvirtualshowcase.o-ran.org/2023/O-RAN_PlugFest_hosted_by_EURECOM_i14y_Lab_Orange_Telefonica_TIM_Vodafone

[10] M. Hoffmann, "How can Energy Saving and Traffic Steering cooperate in O-RAN?", 2024, https://rimedolabs.com/blog/how-can-energy-saving-and-traffic-steering-cooperate-in-o-ran/

[11] Arda Akman, "Enabling Multi-Vendor O-RAN RIC xApps/rApps Coordination with Juniper RAN Intelligent Controller (RIC)", 2024, https://blogs.juniper.net/en-us/driven-by-experience/enabling-multi-vendor-o-ran-ric-xapps-rapps-coordination-with-juniper-ran-intelligent-controller-ric

[12] RIC Forum, March 26–28, 2024 - ITS ) https://its.ntia.gov/research/5g/2024-ric-forum